# Relativistic Proton Production During the 14 July 2000 Solar Event: The Case for Multiple Source Mechanisms

D. J. Bombardieri[1], M. L. Duldig[2], K. J. Michael[1, 4] and J. E. Humble[3, 2]

[1]*Institute for Antarctic and Southern Ocean Studies, University of Tasmania, Hobart, Tasmania, Australia 7001*
[2]*Department of the Environment and Heritage, Australian Antarctic Division, Kingston, Tasmania, Australia 7050*
[3]*School of Mathematics and Physics, University of Tasmania, Hobart, Tasmania, Australia 7001*
[4]*Antarctic Climate and Ecosystems Cooperative Research Centre, University of Tasmania, Hobart, Tasmania, Australia 7001*

*e-mail of corresponding author:* Daniel.Bombardieri@aad.gov.au

ABSTRACT

Protons accelerated to relativistic energies by transient solar and interplanetary phenomena caused a ground-level cosmic ray enhancement on 14 July 2000, Bastille Day. Near-Earth spacecraft measured the proton flux directly and ground-based observatories measured the secondary responses to higher energy protons. We have modelled the arrival of these relativistic protons at Earth using a technique which deduces the spectrum, arrival direction and anisotropy of the high-energy protons that produce increased responses in neutron monitors. To investigate the acceleration processes involved we have employed theoretical shock and stochastic acceleration spectral forms in our fits to spacecraft and neutron monitor data. During the rising phase of the event (10:45 UT and 10:50 UT) we find that the spectrum between 140 MeV and 4 GeV is best fitted by a shock acceleration spectrum. In contrast, the spectrum at the peak (10:55 UT and 11:00 UT) and in the declining phase (11:40 UT) is best fitted with a stochastic acceleration spectrum. We propose that at least two acceleration processes were responsible for the production of relativistic protons during the Bastille Day solar event:

(1) protons were accelerated to relativistic energies by a shock, presumably a coronal mass ejection (CME).

(2) protons were also accelerated to relativistic energies by stochastic processes initiated by magnetohydrodynamic (MHD) turbulence.

1. INTRODUCTION

The relationships between flares and coronal mass ejections (CMEs), and their role in accelerating protons to relativistic energies during major solar eruptive episodes, remain a topic of ongoing research and debate. Reames (1999) argued that the dominant mechanism for the production of relativistic protons observed at 1 AU is via a CME-driven shock and not by processes such as magnetic reconnection associated with solar flares. In contrast, Cane et al. (2003) suggested a two-component mechanism for the production of relativistic particles at 1 AU, arguing that protons are accelerated to relativistic energies by flare processes as well as by shock waves driven out into interplanetary space by CMEs. The application of kinematic models (e.g., the flux-rope catastrophe model of Lin & Forbes 2000; see also Lin et al. 2003 for a review of the various models) hints at the possibility that flares and CMEs might be manifestations of the same eruptive process. Therefore, multiple mechanisms must be considered in the production of relativistic particles.

Ground level enhancements (GLEs) are transient increases in the cosmic ray intensity recorded by ground-based neutron monitors, and are the result of powerful solar processes that accelerate protons to relativistic energies. The energy spectra of relativistic protons from these major solar events carry information about the acceleration process. Therefore, such spectra are useful tools for probing the source mechanisms (Reames 1999). To determine the acceleration process we fit analytical spectral forms representing stochastic and shock acceleration mechanisms to spacecraft and ground-based measurements of relativistic proton fluxes covering the energy spectrum from 140 MeV to 4 GeV. The major aim of this study is to investigate the acceleration process/es responsible for the production of relativistic particles during the Bastille Day solar event.

In § 2 we summarize spacecraft and ground-based observations of the Bastille Day solar event. In § 3 we give a description of the multi-station analysis technique (Cramp et al. 1997*a*) used to model the arrival of relativistic particles at 1 AU. In § 4 we give a description of the analytical shock and stochastic acceleration spectral forms used in the generalised non-linear least squares fitting routine. We fit these spectral forms to ground-based and spacecraft observations of particle fluxes to investigate the acceleration process/es. In § 5 we briefly discuss particle transport conditions during the Bastille Day solar event and, in light of our findings, consider the source mechanisms which may have led to relativistic proton production during this solar event.

2. OBSERVATIONS

The Bastille Day X5.8/3B solar flare and associated full halo CME represent the largest of a series of solar transient phenomena which occurred during a period of intense solar activity extending from 10 to 15 July 2000. This period, described as the 'Bastille Day Epoch' by Dryer et al. (2001), produced three X-class flares (including the Bastille Day flare) and two halo CMEs that were observed with the C2/C3 coronagraphs on board the Solar and Heliospheric Observatory (SOHO) spacecraft. The CMEs, associated shocks and magnetic cloud structures caused major disturbances to the interplanetary magnetic field (IMF) and the geomagnetic field (Dryer et al. 2001). The primary source of this activity was NOAA active region 9077 located near the solar meridian N22°, W07° at the time of the Bastille Day flare.

The flare commenced at 10:03 UT, reached its peak at 10:24 UT and ended at 10:43 UT. Klein et al. (2001) reported prominent bright continuum radio emission which was accompanied by a group of intense Type III bursts from microwave to hectometric wavelengths, with a sudden onset near 10:22 UT and a bright phase between 10:30 and 10:40 UT. Reiner et al. (2001) reported that the flare produced very intense, long-duration Type III radio emissions associated with electron acceleration deep in the solar corona. During its propagation through the solar corona and interplanetary medium, the associated CME generated decametric to kilometric Type II radio emissions (Reiner et al. 2001). Share et al. (2001) reported that hard X-ray and γ-ray line emissions were observed by the HXS and GRS detectors on board the Yohkoh spacecraft at 10:20 UT, approximately four minutes before the peak in soft X-ray emission (10:24 UT). Both emissions peaked at 10:27 UT with γ-ray emission lasting until ~10:40 UT. The HEPAD detectors on board the GOES 8 geostationary satellite recorded sudden increases in relativistic protons (430-745 MeV) between 10:30 and 10:35 UT (Fig. 1).

The GLE onset began between 10:30 and 10:35 UT at several stations, with Thule recording an onset at ~10:32 UT in 1-minute data. The largest neutron monitor responses were observed at South Pole and SANAE with respective maxima in 5-minute data of 58.3% and 54.5% above the pre-increase levels (Fig. 2, upper panel). The impulsive nature of the neutron monitor intensity/time profiles (Fig. 2) indicates that relativistic protons had rapid access to Sun-Earth connecting field lines. The event was seen at Climax, indicating the presence of particles with rigidity of at least 3.0 GV. The Lomnicky Stit neutron monitor (with a geomagnetic cut-off of 4.0 GV and not shown in Fig. 2) recorded an increase of marginal significance that may or may not be related to the GLE.

The lower panel of Figure 2 also shows the details of the pressure-corrected intensity-time profiles for the monitors located at Thule and Tixie Bay, while Figure 3 shows the viewing directions of selected neutron monitors at 10:40 UT. We note that Thule, with a viewing direction near the nominal Parker spiral, observed an earlier onset and more rapid rise than did Tixie Bay, whose viewing direction was close to the anti-sunward field direction.

Corrections of observed increases to a standard sea-level atmospheric depth of 1033 g cm$^{-2}$ were made using the two-attenuation length method of McCracken (1962). The attenuation length for solar cosmic rays can be determined from the ratio of increases at two stations with similar viewing directions but different altitudes (Cramp et al. 1997 *b*). An attenuation length of 110 g cm$^{-2}$ was derived from a comparison of data from Mt Wellington, Hobart and Kingston neutron monitors.

In order to calculate the absolute flux of solar particles, it is necessary to select a low altitude station with one of the largest increases as the normalization station (Cramp 1996). After correcting observed increases to standard sea-level atmospheric depth, SANAE was found to have the largest response and was used as the normalization station for this analysis.

3. MODELLING THE NEUTRON MONITOR RESPONSE

The technique for modelling the solar cosmic ray ground level response by neutron monitors has been developed over many years (Shea & Smart 1982; Humble et al. 1991) and is described in detail in Cramp et al. (1997*a*). The geomagnetic field model of Tsyganenko (1989), with IGRF 2000 parameters and adjustments for Kp and the DST index, was employed to determine the asymptotic viewing directions of ground-based instruments (Flückiger & Köbel 1990; Boberg et al. 1995). The successful use of this technique depends on collecting data from many stations widely separated in both latitude and longitude. A range of cut-off rigidities (geomagnetic latitudes) allows the determination of spectral characteristics, whilst a range of latitudes and longitudes are necessary to determine the extent of anisotropy.

To better resolve the responses of neutron monitors and produce a more accurate model of the arrival of relativistic particles at the Earth, we use asymptotic viewing cones (the set of asymptotic directions of all allowed trajectories) calculated at nine different arrival directions (vertical; and 90°, 180°, 270° and 360° azimuth at 16° and 32° zenith). The increasing solid angle away from the zenith compensates for the decreasing flux caused by increased atmospheric attenuation; therefore each cone represents an approximately equal contribution to the total counting rate (Rao et al. 1963; see Cramp et al. 1997*b* for a complete review). Our trajectory calculations are therefore a more accurate representation of the asymptotic cone of view compared to models using simple vertical approximation methods (e.g., Belov et al. 2001; Bieber et al. 2002; Vashenyuk et al. 2003).

A least-squares fitting technique, minimizing the difference between the observed response corrected to sea level and the equivalent calculated response for each neutron monitor, was used to determine the axis of symmetry of particle arrival, particle pitch angle distribution, and rigidity spectrum. The model should accurately reproduce the observed increases as well as produce null responses for those stations that did not record an intensity increase. Inclusion of data from stations with null responses places additional bounds on the spectra and anisotropy characteristics. Furthermore, the least-squares fitting technique allows us to efficiently analyse parameter space and derive an optimal solution for each of the time intervals considered.

3.1. *Results*

Data from 30 neutron monitors (Table 1) were modelled every five minutes between 10:35 and 10:55 UT during the rise and peak phases of the event. During the decay phase, data were modelled every ten minutes from 11:00 to 15:00 UT. Each indicated time represents the start of a five-minute integrated time interval. Parameter determinations are less accurate later in the event, when the increase above background is small. Fits were discontinued at 15:00 UT when the increase above the background at the normalization station (SANAE) was small (~10 %).

Figure 4 shows the observed increases, corrected to standard sea level pressure, at selected neutron monitor stations (representing a range of geomagnetic cut-offs from 0.01 - 3.03 GV) and the model fits to those observations. Good fits to observations were achieved during all phases of the Bastille Day GLE. However, South Pole and Mawson responses were not as well fitted during the high intensity phase of the event. In particular, the model slightly overestimated the neutron monitor response at Mawson and underestimated the neutron monitor response at South Pole

### 3.1.1. *Arrival Directions*

Figure 5 illustrates the GSE latitude and longitude of the apparent arrival directions, together with the IMF direction as measured by ACE. The method of Bieber et al. (2002) was replicated to permit direct comparisons with ACE measurements and their results.

The average GSE longitude of the IMF direction as measured by ACE was 330º, which implies that particles were flowing from the Sun close to a nominal Parker spiral. We find the apparent longitude of the arrival direction between 10:30 and 11:00 UT to be centred slightly east of the Sun-Earth line. This result is ~ 30º east of the measured field direction and 15º east of that calculated by Bieber et al. (2002). The difference might be due to our method (utilizing higher cut-off rigidity stations) probing the IMF at larger scales than does the method of Bieber et al. (2002) which utilizes only high latitude stations with low cut-off rigidities (private communication, J. Bieber, Bartol Research Institute, 2005). Between 11:10 and 12:40 UT both our model and that of Bieber et al. (2002) show good agreement with the measured IMF longitude. However, from 12:40 to 15:00 UT our model longitudes move east of the measured field longitude by up to ~120º. The results of Bieber et al. (2002) for the same interval also show poor agreement with the measured field longitude.

From 10:30 to 10:55 UT the apparent latitude of the arrival direction shows good agreement with the measured field latitude. However, during the decline phase, our model latitudes, and those of Bieber et al. (2002), are in poor agreement with the measured field latitude, although they do follow the overall southerly trend.

As noted by Bieber et al. (2002) there is no reason why the magnetic field measured at a point should be the same as the average field sampled by the particle over its orbit. For example, a 2 GV proton has a Larmor radius of ~0.01 AU which is of the order of the coherence length of interplanetary magnetic turbulence. Therefore, model flow vectors need not align exactly with the measured magnetic field vector.

3.1.2. *Particle Anisotropy*

The particle pitch angle $\alpha$ ($\theta, \phi$) is the angle between the axis of symmetry of the particle distribution ($\theta_s, \phi_s$) and the asymptotic direction of view at rigidity $P$ associated with the arrival direction ($\theta, \phi$). The pitch angle distribution is a simplification of the exponential form described by Beeck & Wibberenz (1986). It has the functional form

$$G(\alpha) = exp\left[\frac{-0.5(\alpha - \sin\alpha \, \cos\alpha)}{A - 0.5(A - B)(1 - \cos\alpha)}\right] \qquad (1)$$

where $A$ and $B$ are variable parameters (Cramp et al. 1997*a*). The temporal development of the pitch angle distribution during the Bastille Day GLE is presented in Figure 6. Near onset (10:35 UT), the particle arrival was strongly anisotropic. The anisotropy decreased rapidly over the next 20 minutes and remained relatively unchanged thereafter. The particle distribution can be divided into an anisotropic component representing particles which arrive directly from the Sun and an isotropic component where the effects of local scattering dominated the distribution. Local scattering effects increased as the event progressed.

3.1.3. *Spectrum*

One advantage of our modelling technique is the ability to utilize various spectral forms such as pure and modified power laws, as well as theoretical shock acceleration spectra (Ellison & Ramaty, 1985), to achieve the best fit between observed and calculated responses. In contrast, Bieber et al. (2002) determined spectral exponents from the ratio of count rates of the standard (NM64) neutron monitor at the South Pole and an unshielded (Polar Bare) neutron monitor at the same site. Their technique utilizes the different response functions of the neutron monitors and is independent of particle anisotropy. The yield function used by Bieber et al. (2002) differs from that used in our model (see Cramp et al. 1997*a* for a complete review). Our model generally produces steeper power law spectra than the model of Bieber & Evenson (1991) and Bieber et al. (2002). Similarly Lockwood et al. (2002), employing the same yield function as our model but using the same method to calculate spectral exponents as Bieber & Evenson (1991), albeit with different stations (e.g., Mt. Washington and Durham), also obtain steeper spectral exponents.

For this part of the analysis we used a simple power law as well as a modified power law with more rapidly steepening slope (equation 2)

$$J_{\parallel} = KP^{(\gamma-\delta\gamma(P-1))} \qquad (2)$$

where $J_{\parallel}$ is the peak cosmic ray flux arriving from the Sun along the axis of symmetry of the pitch angle distribution. The parameters are the particle rigidity ($P$), the parallel flux at 1 GV ($K$), the power law exponent ($\gamma$) and the change of $\gamma$ per GV ($\delta\gamma$), where a positive value of $\delta\gamma$ results in a spectrum that steepens with increasing rigidity.

Finally, we used an approximation of the Ellison & Ramaty shock spectrum as described in Cramp (1996), namely a power law with exponent $-\gamma-(1-\beta^2)(1+\delta\gamma)$. Here $\gamma$ is the spectral index, $\beta$ is the ratio of the particle speed to the speed of light and $\delta\gamma$ is an exponent modifier to account for a non-infinite shock interacting for a finite time.

Our modelling showed that the modified power law spectral form, in general, produced the best fit during all phases of the event (Table 2). The derived particle spectra are illustrated in Figure 7. The spectral slope varied considerably during the rise phase of the event (10:35-10:55 UT). At 10:35 UT the spectrum was represented by a power law but by 10:40 and 10:45 UT the change of slope parameter (δγ) was significantly greater than zero. By 11:00 UT the spectrum again had small values of δγ. Furthermore, we found that the best fit spectra in the present event were insensitive to anything other than gross changes in the arrival direction or pitch angle distribution

*3.2. Confidence Limits on Model Parameters*

Rigorous error analyses of the derived parameters are difficult due to the complexity of the model and the strong interdependence between the parameters of the fit. An attempt to estimate the uncertainty of the derived parameters can be made by considering the relative changes in the sum of squares between the observed and calculated increases for each solution, giving a measure of the significance of the change in the parameters.

Uncertainties for the geographic latitude and longitude of the apparent arrival directions are influenced by the adequacy of the asymptotic direction calculations to describe the actual propagation of the particles through the magnetosphere (Lovell et al. 1998). The degree of anisotropy of the particle distribution is also an important factor. Broader pitch angle distributions result in less confidence in the axis of symmetry. We estimate the uncertainty for the apparent particle arrival directions at 10:35 UT to be ±8° in latitude and ±16° in longitude. At 13:25 UT these uncertainties are estimated to be ±20° in both latitude and longitude. Uncertainties for parameters at most other solutions will lie between these values.

The uncertainty of the spectral slope (γ) at 10:35 UT is expected to exceed that at most other times due to the dominance of the particle anisotropy at this time. The very small spatial extent of the particle arrival distribution means that only a few stations with similar asymptotic viewing directions and rigidity apertures observed this part of the event. Consequently, spectral information is restricted to a narrow rigidity range leading to an uncertainty in the slope. The spectral slope (γ) at 10:35 UT is -5.2 ±1.0. At 13:25 UT γ is -8.0 ±0.1. The uncertainty in the change of slope (δγ) at 10:35 is small, while at 13:25 UT the uncertainty in δγ is estimated at ± 0.2. The resulting uncertainty in the calculated flux at 1 GV is less than 10%.

4.0. PARTICLE ACCELERATION SPECTRA

Various mechanisms have been proposed for the acceleration of particles during extreme solar events. These include direct acceleration by DC electric fields in neutral current sheets, diffusive shock acceleration at the bow shock of a CME and resonant wave-particle interactions (stochastic acceleration) initiated by MHD turbulence (e.g., Miller et al. 1997). In this section we briefly discuss the analytical spectra used in our modelling.

4.1. *Diffusive Shock Acceleration*

Ellison & Ramaty (1985) derived an equation for spectra resulting from diffusive shock acceleration. Particles are able to gain energy by scattering multiple times between magnetic field irregularities both upstream and downstream of the shock. The compression at the shock is the source of the energy. The differential particle intensity *dJ/dE* is given by a power law spectrum truncated with an exponential (see equation 3). A variety of effects may explain the exponential roll-off at higher energies; for example, particle acceleration is less effective above the energy $E_0$ (*e*-folding energy) where proton intensities can no longer sustain the growth of resonant waves. High-energy particles begin to leak from the acceleration region, truncating the power law behaviour (Reames 2000)

$$\left(\frac{dJ}{dE}\right) \propto \left(\frac{dJ}{dE}\right)_0 \exp\left(-\frac{E}{E_0}\right) \qquad (3)$$

where,

$$\left(\frac{dJ}{dE}\right)_0 \propto n_{inj}\left(E_i^2 + 2E_i m_0 c^2\right)^{3/[2(r-1)]}\left(E^2 + 2E m_0 c^2\right)^{-(1/2)[(r+2)/(r-1)]} \qquad (4)$$

($dJ/dE$)$_0$ is the differential particle intensity (particle/cm$^2$ s sr MeV), $n_{inj}$ is the number density of seed particles injected far upstream of the shock, $c$ is the speed of light, $m_o c^2$ is the proton rest mass energy, $E$ is the particle energy in MeV (Ellison & Ramaty 1985). The variable parameters of equation (4) are the shock compression ratio $r$ (the ratio of the upstream and downstream flow velocities) and the $e$-folding energy $E_0$ (MeV), with the spectral index being a function of $\sigma = 3r/(r-1)$. For the sake of clarity, the final exponent in equation (4) differs from its appearance in equation (1) of Ellison & Ramaty (1985), where it was capable of ambiguous interpretation.

## 4.2. *MHD Turbulence*

Energy from MHD turbulence is transferred to particles through the process of wave-particle resonant interactions. The origin of this turbulence is still uncertain and a topic of much conjecture. However, it is thought that turbulence in the form of Alfvèn waves is generated at large wavelengths by plasma outflow jets created at magnetic reconnection sites or by large-scale magnetic field perturbations (Miller et al. 1997; Priest & Forbes 2002).

Results of a study by Galsgaard et al. (2005), involving three-dimensional numerical simulations of photospheric flux emergence with a simple coronal field configuration, reveal the formation of arched high-density current sheets. Sites of magnetic reconnection within the current sheets produced high-velocity plasma outflow jets. These jets could represent the source of the MHD turbulence which initiates stochastic acceleration. With the advent of the RHESSI spacecraft there is now strong observational evidence supporting the importance of current sheets in major solar eruptive episodes (Ciaravella et al. 2002; Ko et al. 2003; Sui & Holman 2003; Webb et al. 2003; Gary & Moore 2004; Sui et al. 2004, Lin et al. 2005).

Stochastic acceleration can be described by a Fokker-Plank equation in energy space. This equation accounts for the diffusive and convective nature of the process. Perez-Pereza & Gallegos-Cruz (1994) and Gallegos-Cruz & Perez-Pereza (1995) presented solutions to the Fokker-Plank equation in energy space, based on the WKBJ approximation method (see above authors for a complete review of the method). These solutions are valid over the entire energy range (i.e., non-relativistic, trans-relativistic, ultra-relativistic), for both time-dependent and steady-state conditions. For this study, we used the steady-state energy spectrum for MHD turbulence assuming mono-energetic injection (see equation (5), derived from equation (43) in Gallegos-Cruz & Perez-Pereza 1995).

$$N(E) \cong \frac{(q_0/2)(\beta_0/\beta)^{1/4}(\varepsilon/\varepsilon_0)^{1/2}}{(\alpha/3)^{1/2} a^{1/4}(E) a^{1/4}(E_0) \beta_0^{3/2} \varepsilon_0} \left[ \frac{\varepsilon + \beta\varepsilon}{\varepsilon_0 + \beta_0\varepsilon_0} \right]^{-(b+1)/2b}$$
$$\times \exp\left[ \left(\frac{-1}{2b}\right)\left(\beta^{-1} - \beta_0^{-1}\right) \right]$$

(5)

$N(E)$ = Particles per unit energy, $q_0$ = rate of particle injection (s$^{-1}$), $\beta = v/c$, $E_0$ = energy of injected particles (MeV), $a \approx (\delta + \alpha)\beta + (\alpha/3)\beta^{-1}$, where $\delta$ is set to 1 s$^{-1}$, $\varepsilon$ = energy + proton rest mass energy, $b = [(3/\alpha)(\delta + \alpha)]^{1/2}$. The variable parameters for equation (5) are the acceleration efficiency $\alpha$ and normalization factor $N$. Again, for the sake of clarity, the exponent in equation (5) differs from its appearance in equation (43) of Gallegos-Cruz & Perez-Pereza 1995).

The analytical spectra deduced from the neutron monitors were used to generate the input to the fitting routine at 10 selected energies spaced evenly on a logarithmic scale. The fitting routine was a generalised non-linear least squares program with data points for spacecraft and neutron monitor energy ranges weighted by errors in the flux data. Due to the difficulties in conducting absolute flux measurements in orbit, there is no quantitative knowledge of the errors/confidence limits for GOES 8 spacecraft particle data. However, estimates set these errors to within a factor of two of the measured value (private communication, T. Onsager, GOES PI, 2004).

*4.3. Results*

Tables 3 and 4 present the results of the spectral fits to combined spacecraft and neutron monitor data at times 10:45 UT, 10:50 UT, 10:55 UT, 11:00 UT and 11:40 UT. It should be noted that the full Ellison & Ramaty spectral form (equations 3 & 4) was used during this stage of the analysis. Spectral fits to combined spacecraft and neutron monitor data during the early rise phase of the event were not successful due to the inconsistency of spacecraft and ground-based intensity measurements at 1 GV.

Figure 8 illustrates the results of fitting the analytical shock and stochastic acceleration spectra to relativistic proton fluxes determined from spacecraft and neutron monitor observations at 1 AU. A Kolmogorov-Smirnov test at 95% confidence shows that all post fit residuals were random, giving confidence in the weighted sum of squares result. For comparison with spacecraft results, Figure 8 shows energy rather than rigidity spectra. Note in Figure 8*a*, and to a lesser degree in Figure 8*b*, the turnover in the low energy spectra. It appears that the effect of velocity dispersion for a 140 MeV particle was significant for this event, particularly at 10:45 UT. As a result, for this interval, we also fitted the analytical shock and stochastic acceleration spectral forms to relativistic proton fluxes from ~300 to 4000 MeV rather than 140 to 4000 MeV as used at other times.

Table 3 lists the results and standard errors for the variable model parameters (compression ratio and *e*-folding energy $E_0$) from the shock acceleration non-linear least squares fitting routine. The proton spectrum at 10:45 UT and 10:50 UT is best fitted with this spectral form. The shock compression ratios for these intervals are 1.95 ± 0.03 and 1.83 ± 0.02 respectively with *e*-folding energies of 1.87 GeV ± 0.01 and 1.94 ± 0.03 GeV respectively. The value of the *e*-folding energy at 10:50 UT (2.8 GV in terms of rigidity) is consistent with the maximum proton rigidity of ~3 GV observed for this event.

The spectra at the peak (10:55 and 11:00 UT) and in the declining phase (11:40 UT) are best fitted by a stochastic acceleration spectral form; implying acceleration via resonant wave-particle interactions initiated by MHD turbulence. Table 4 lists the results and standard errors for the variable model parameters (normalization factor $N$ and acceleration efficiency $\alpha$) from the stochastic non-linear least squares fitting routine. For the time intervals we modelled, $\alpha$ ranged from 0.03 - 0.05 s$^{-1}$. This implies that, to produce the observed response, protons with injection energy $E_0$ of 1 MeV need only a modest acceleration efficiency, which is consistent with values reported from previous studies (Murphy and Ramaty 1984; Miller et al. 1990; Miller 1991).

5.0. DISCUSSION
5.1. *Particle Scattering*

Particle pitch angle distributions (Fig. 6) provide information on the homogeneity of the interplanetary magnetic field. The particle arrival near GLE onset (10:35 UT) was clearly anisotropic, indicating focused transport conditions. This relatively strong anisotropy decreased rapidly over the next 20 minutes indicating that the protons experienced significant scattering. Bieber et al. (2002) proposed that the rapid decrease in anisotropy for this event was strongly influenced by a magnetic mirror located 0.3 AU beyond the Earth, which reflected ~ 85% of the relativistic solar protons back toward the Earth. Their hypothesis is supported by ACE and WIND spacecraft observations of shocks and associated magnetic structures which passed the Earth on 13 July, as well as the rapid increase in the neutron monitor response of stations viewing in the anti-sunward field direction (e.g., Tixie Bay).

Magnetic mirroring, as evidenced by bi-directional flow of relativistic particles following intense solar activity, has been previously reported by Cramp et al. (1997*a*). We examined the likelihood of bi-directional flow by using the GLE modelling technique of Cramp et al. (1997*a* and 1997*b*), incorporating a modification of the pitch angle distribution function of equation (1) as follows:

$$G'(\alpha) = G_1(\alpha) + C \times G_2(\alpha'), \tag{6}$$

where $G_1$ and $G_2$ are of the same form as in equation (1) with independent parameters $A_1$, $B_1$, $A_2$ and $B_2$; $\alpha' = \pi - \alpha$; and $C$ is the ratio of reverse-to-forward flux ranging from 0 and 1. Figure 9 illustrates the fits of the bi-directional pitch angle distribution functions to the data. We do not observe an excess of reverse-propagating particles (i.e., a significant peak in the pitch angle distribution centred at 180°). Furthermore significant increases in neutron monitor responses at 10:40 UT (approximately 10 minutes after GLE onset) were not only observed at stations viewing in the anti-sunward field direction (e.g., Tixie Bay, 12.8%), but at stations viewing perpendicular from the nominal sunward field direction (Apatity, 30.6% and Inuvik, 15.3%). Therefore, we conclude that the underlying isotropic component (Fig. 6) was due to local scattering effects such as magnetic field turbulence.

*5.2. Source Mechanisms*

The major finding of this study is that during the Bastille Day 2000 solar event at least two distinct acceleration processes (shock and stochastic) operated to produce relativistic protons (Fig. 8). During the rise phase of the event (10:45 UT and 10:50 UT) the best-fit spectral form is shock acceleration (Tables 3 and 4). At the peak (10:55 UT, 11:00 UT) and during the declining phase (11:40 UT) the best-fit spectral form is clearly stochastic acceleration.

Perez-Peraza et al. (2003), investigating the origin of relativistic protons for the Bastille Day solar eruptive episode, also showed that the process of stochastic acceleration was important in relativistic particle production at 1 AU. They proposed that relativistic protons were stochastically accelerated by MHD turbulence associated with a flare-generated expanding closed magnetic structure in the low corona. These particles were then injected into interplanetary space either as a consequence of the opening of the closed magnetic structure due to plasma instabilities, or perhaps they were carried into interplanetary space by an expanding CME.

In addition, Klein et al. (2001) using radio, X-ray, EUV and visible light observations, traced the non-radial propagation path of a filament to the northwestern solar quadrant. They proposed that this filament interacted with coronal structures (large-scale coronal loops) near to Sun–Earth connecting magnetic field lines (i.e., near 60° western heliolongitude). This interaction involved reconfiguration of the coronal magnetic field in the wake of the erupting filament (CME). Klein et al. (2001) based this finding on radio observations of a prominent bright continuum radio source, accompanied by a group of intense Type III radio bursts from microwave to hectometric wavelengths, which coincided with a rise in neutron monitor count rates. They propose that reconfiguration of the coronal magnetic field led to relativistic proton production and that the major driver of these changes was the ejected magnetic field configuration around the erupting filament which was part of the CME

Our results show that at 10:45 to 10:50 UT during the Bastille Day 2000 solar eruptive event a shock was responsible for the production and arrival of relativistic protons at 1 AU. This is supported by the detection of Type II decametric to kilometric radio emissions as the shock propagated through the corona (Reiner et al. 2001). The most likely source for this shock other than a flare or coronal source was the Bastille Day CME. Furthermore, our modelling shows that the spectral form changed at 10:55 UT implying that the source of relativistic protons also changed. This new source may be attributed to magnetic reconnection sites which produced high velocity plasma outflow jets as a result of reconfiguration of the coronal magnetic field in the wake of the CME. These jets may represent the source of the MHD turbulence that initiated stochastic acceleration.

Our findings in part support those of Klein et al. (2001) which suggest that reconfiguration of the coronal magnetic field led to relativistic proton production and that the bow shock of the Bastille Day CME was not the sole accelerator of relativistic particles for this event.

6. CONCLUSION

We have modelled the arrival of relativistic protons at 1 AU for the Bastille Day 2000 solar eruptive episode. The GLE was an impulsive event as shown by the neutron monitor intensity/time profiles. This suggests that relativistic protons had rapid access to Sun-Earth connected magnetic field lines. We find that the event was marked by a highly anisotropic onset followed by a rapid decrease in anisotropy, and attribute this result to the effects of turbulence associated with the interplanetary magnetic field. Our modelling also shows that the spectrum varied considerably during the rising phase of the event.

We employed theoretical shock and stochastic acceleration spectral forms in our fits to spacecraft and neutron monitor data over the energy range 140 MeV to 4 GeV to investigate the acceleration process. We found the spectrum during the rise phase (i.e., at 10:45 and 10:50 UT) was best fitted with a shock acceleration spectral form, implying acceleration of protons to relativistic energies at a coronal shock or at the bow shock of the Bastille Day CME. In contrast, the spectrum at the peak and declining phase (i.e., at 10:55 and 11:40 UT) was best fitted with a stochastic acceleration spectral form, implying acceleration of protons to relativistic energies by stochastic processes via MHD turbulence. The change in spectral form represents a new source of relativistic particle production other than a shock. We propose this source to be magnetic reconnection sites created by the reconfiguration of the coronal magnetic field in the wake of the CME.

ACKNOWLEGMENTS

We thank our colleagues at IZMIRAN (Russia) and The Polar Geophysical Institute (Russia) for contributing neutron monitor data. D. J. Bombardieri acknowledges receipt of an Australian Postgraduate Award and Australian Antarctic Science Scholarship as well as support from the University of Tasmania. Neutron monitors of the Bartol Research Institute are supported by NSF ATM 00-00315. GOES 8 data were obtained from Space Physics Interactive Data Resource http://spidr.ngdc.noaa.gov/.

REFERENCES

Beeck, J., & Wibberenz, G. 1986, ApJ, 311, 437

Belov, A. V., et al. 2001, Proc. 27th Int.Cosmic-Ray Conf. (Hamburg), 8, 3446

Bieber, J. W., & Evenson, P. 1991, Proc. 22nd Int.Cosmic-Ray Conf. (Dublin), 3, 129

Bieber, J. W., et al. 2002 ApJ, 567, 622

Boberg, P. R., Tylka, A. J., Adams, Jr., J. H., Flückiger, E. O., & Kobel, E. 1995, Geophys. Res. Lett., 22, 1133

Cane, H. V., von Rosenvinge, T. T., Cohen, C. M. S., & Mewaldt, R. A. 2003, Geophys. Res. Lett., 30, SEP 5-1

Ciaravella, A., Raymond, J. C., Li, J., Reiser, P., Gardner, L. D., Ko, Y. –K., & Fineschi, S. 2002, ApJ, 575, 1116

Cramp, J. L. 1996, PhD Thesis

Cramp, J. L., Duldig, M. L., Flückiger, E. O., & Humble, J. E. 1997a, J. Geophys. Res., 102, 24237

Cramp, J. L., Duldig, M. L., & Humble, J. E. 1997b, J. Geophys. Res., 102, 4919

Dryer, M., Fry, D. D., Sun, W., Deehr, C., Smith, Z., Akasofu, S. –I., & Andrews, M. D. 2001, Space Sci., 204, 267

Ellison, D. C., & Ramaty, R. 1985, ApJ, 298, 400

Flückiger, E., & Köbel, E. 1990, J. Geomag. Geoelect., 42, 1123

Gallegos-Cruz, A., & Perez-Pereza, J. 1995, ApJ, 446, 400

Galsgaard, K., Moreno-Insertis, F., Archontis, V., & Hood, A. 2005, ApJ, 618, L156

Gary, A. J., & Moore, R. L. 2004, ApJ, 611, 545

Humble, J. M., Duldig, M. L., Smart, D., and Shea, M. 1991, Geophys. Res. Lett., 18, 737

Ko, Y.-K., Raymond, J. C., Lin, J., Lawrence, G., Li, J., & Fludra, A. 2003, ApJ, 594, 1068

Klein, K.L., Torttet, G., Lantos, P., & Delaboundiniere, J.-P. 2001, A&A, 373, 1073

Lin, J., & Forbes, T. G. 2000, J. Geophys. Res., 105, 2375

Lin, J., Soon, W., & Baliunas, S. L. 2003, NewA. Rev., 47, 53

Lin, J., et al. 2005, ApJ, 622, 1251

Lockwood, J. A., Debrunner, H., Flueckiger, E. O., Ryan, J. M. 2002, Sol. Phys., 208, 113

Lovell, J. L., Duldig, M. L., & Humble, J. E. 1998, J. Geophys. Res., 103, 23733

McCracken, K. G. 1962, J. Geophys. Res., 67, 423

Miller, J. A., Guessoum, N., & Ramaty, R. 1990, ApJ, 361, 701

Miller, J. A. 1991, ApJ, 376, 342

Miller, J. A., et al. 1997, J. Geophys. Res., 102, 14631

Murphy, R. J., & Ramaty, R. 1984, Adv. Space Res., 4, No. 7, 127

Perez-Pereza, J., & Gallegos-Cruz, A. 1994, ApJS, 90, 669

Perez-Peraza, J, Gallegos-Cruz, A., Vashenyuk, E. V., & Miroshnichenko, L. I. 2003, Proc. 28th Int.Cosmic-Ray Conf. (Tsukuba), 6, 3327

Priest, E. R., & Forbes, T. G. 2002, Astron. Astrophys. Rev., 10, 313

Rao, U. R., McCracken, K. G., & Venkatesan D. 1963, J. Geophys. Res., 68, 345

Reames, D, V. 1999, Space Sci. Rev., 90, 413

Reames, D. V. 2000, in AIP Conf. Proc., 528, Acceleration and Transport of Energetic Particles Observed in the Heliosphere, eds. R.A. Mewaldt, J.R. Jokipii, M.A. Lee, E. Moebius, & T.H. Zurbuchen, (New York: AIP), 79

Reiner, M. J., Kaiser, M. L., Karlicky, M., Jiricka, K., & Bougeret J. –L. 2001, Space Sci., 204, 123

Share, G.H., Murphy, R. J., Tylka, A.J., Schwartz, R. A., Yoshimori, M., Suga, K., Nakayama, S., & Takeda, H. 2001, Sol. Phys., 204, 43

Shea, M. A., & Smart, D. F. 1982, Space Sci. Rev., 32, 251

Sui, L., & Holman, G. D. 2003, ApJ, 596, L251

Sui, L., Holman, G.D., & Dennis, B.R. 2004, ApJ, 612, 546

Tsyganenko, N. A. 1989, Planet. Sp. Sci., 37, 5

Vashenyuk, E. V., Balabin, Y. V., & Gvozdevsky, B. B. 2003, Proc. 28th Int. Cosmic-Ray Conf. (Tsukuba), 6, 3401

Webb, D. F., Burkepile, J., Forbes, T.G., & Riley, P. 2003, J. Geophys. Res., 108, 1440

FIGURE CAPTIONS

FIG. 1 ------ Five-minute GOES 8 observations of proton fluxes associated with the Bastille Day 2000 solar event. P6 to P10 represent the EPS/HEPAD sensor/detector differential energy channels (particles/cm$^2$.s.sr.MeV) with the following characteristics of nominal energy range (MeV) and midpoint energy (MeV; brackets): P6 = 84-200 (142); P7 = 110-500 (305); P8 = 370-480 (425); P9 = 480-640 (560); P10 = 640-850 (745).

FIG. 2. ------ Solar cosmic ray intensity/time profiles for 14 July 2000 as recorded by South Pole and SANAE neutron monitors (*top*), and Thule and Tixie Bay neutron monitors (*bottom*). The viewing directions of the Thule and Tixie Bay neutron monitors approximately represent the sunward and anti-sunward field direction, respectively. The impulsive nature of the neutron monitor intensity/time profiles (i.e., fast rise to maximum) is typical of well-connected events.

FIG. 3. ------ Viewing directions of neutron monitors in geographic coordinates at 10:40 UT (10 minutes after GLE onset) on Bastille Day 2000. Geomagnetic conditions were slightly disturbed (Kp = 4; DST = -18). Lines for each station represent the vertical viewing direction at different rigidities. Numeral 4 represents the vertical viewing direction at maximum rigidity (~4 GV), while numeral 1 represents the vertical viewing direction at the atmospheric cutoff (~1 GV). The solid circles show the median rigidity of response to the GLE for each station. O and X designate the position of the nominal sunward and anti-sunward field direction respectively. Station abbreviations are: APT = Apatity, Russia; GSB = Goose Bay, Canada; IVK = Inuvik, Canada; KIN = Kingston, Australia; MAW = Mawson, Antarctica; MCM = McMurdo, Antarctica; SAN = SANAE, Antarctica; SPO = South Pole, Antarctica; TER = Terre Adelie, Antarctica; THU = Thule, Greenland; TXB = Tixie Bay, Russia.

FIG. 4. ------ The observed (line) and modelled (solid circles) responses to the Bastille Day 2000 GLE. The selected neutron monitor stations represent a range of vertical geomagnetic cut-offs ($P_c$: 0.1 to 3.0 GV).

FIG. 5. ------ GSE longitude (*top*) and GSE latitude (*bottom*) of the apparent arrival directions (this study; solid circles) plotted with the negative magnetic field direction (1 hour centred moving averages; line) as measured by the MAG instrument onboard the ACE spacecraft.

FIG. 6. ------ Derived pitch angle distributions for 10:35 UT, 10:40 UT, 10:45 UT (*top*) and for 10:50 UT, 10:55 UT and 11:00 UT (*bottom*).

FIG. 7. ------ Derived rigidity spectra for 10:35 UT, 10:40 UT, 10:45 UT (*top*) and for 10:55 UT, 11:00 UT and 11:40 UT (*bottom*).

FIG. 8. ------ Energy spectral fits to combined satellite and ground based observations. Five-minute proton data (*solid circles*) from GOES 8 EPS/HEPAD particle detectors; energy range is ~100 to 700 MeV. The neutron monitor derived data (*open circles*) range from ~400 to 4000 MeV and are spaced evenly on a logarithmic scale. Fitted curves are of the Ellison & Ramaty (1985) shock acceleration (*line*) and Gallegos-Cruz & Perez-Pereza (1995) stochastic acceleration (*dashed line*) spectral forms: (*a*) 10:45 UT rise phase (140 to 4000 MeV); (*b*) 11:00 UT peak phase; (*c*) 11:40 UT decline phase for the Bastille Day 2000 GLE.

FIG. 9. ------Derived pitch angle distributions incorporating bi-directional flow parameters for 10:35 UT, 10:40 UT, 10:45 UT (*top*) and 10:50 UT, 10:55 UT and 11:00 UT (*bottom*).

# Table 1

TABLE I

NEUTRON MONITORS AND GEOMAGNETIC CUT-OFF RIGIDITIES

| Station | Lat. (deg.) | Lon. (deg.) | $P_c^a$ (GV) | Alt. (m) |
|---|---|---|---|---|
| Apatity | 67.55 | 33.33 | 0.61 | 177 |
| Aragats | 40.50 | 44.17 | 7.60 | 3200 |
| Climax | 39.37 | 253.82 | 3.03 | 3400 |
| Goose Bay | 53.27 | 299.60 | 0.52 | 46 |
| Haleakala | 20.27 | 203.73 | 13.3 | 3033 |
| Hermanus | -34.42 | 19.22 | 4.90 | 26 |
| Hobart | -42.90 | 147.33 | 1.88 | 18 |
| Inuvik | 68.35 | 226.28 | 0.18 | 21 |
| Jungfraujoch | 46.55 | 7.98 | 4.48 | 3475 |
| Kerguelen Island | -49.35 | 70.25 | 1.19 | 33 |
| Kiel | 54.33 | 10.13 | 2.29 | 54 |
| Kingston | -42.99 | 147.29 | 1.88 | 65 |
| Larc | -62.20 | 301.04 | 2.21 | 40 |
| Lomnicky Stit | 49.20 | 20.22 | 4.00 | 2634 |
| Magadan | 60.12 | 151.02 | 2.10 | 220 |
| Mawson | -67.60 | 62.88 | 0.22 | 30 |
| McMurdo | -77.85 | 166.72 | 0.01 | 48 |
| Moscow | 55.47 | 37.32 | 2.46 | 200 |
| Mt. Wellington | -42.92 | 147.23 | 1.89 | 725 |
| Newark | 39.68 | 284.25 | 1.97 | 50 |
| Oulu | 65.05 | 25.47 | 0.81 | 15 |
| Potchefstroom | -26.68 | 27.10 | 7.30 | 1351 |
| Rome | 41.86 | 12.47 | 6.32 | 0 |
| Sanae | -71.67 | 357.15 | 1.06 | 856 |
| South Pole | -90.00 | 0.00 | 0.10 | 2820 |
| Terre Adelie | -66.67 | 140.02 | 0.01 | 45 |
| Thule | 76.50 | 291.30 | 0.00 | 260 |
| Tixie Bay | 71.58 | 128.92 | 0.53 | 0 |
| Tsumeb | -19.20 | 17.58 | 9.29 | 1240 |
| Yakutsk | 62.03 | 129.73 | 1.70 | 105 |

[a] Vertical geomagnetic cutoff rigidities represent the minium rigidities below which particles do not have access to a particular site on the Earth's surface. The cut-off at the geomagnetic equator is ~ 17 GV, decreasing to zero at the geomagnetic poles.

# Table 2

TABLE II

MODEL PARAMETERS AND ASSOCIATED SPECTRAL FORMS

| | | Power Law | | | Modified Power Law | | | | Modified Ellison & Ramaty | | | |
|---|---|---|---|---|---|---|---|---|---|---|---|---|
| Time[a] | Inc[b] | $J_\parallel$[c] | $\gamma$[d] | wss[e] | $J_\parallel$ | $\gamma$ | $\delta\gamma$[f] | wss[g] | $J_\parallel$ | $\gamma$ | $\delta\gamma$[h] | wss[i] |
| 10:30 | 3.39  | 2   | -4.35 | 22  | 2   | -4.30  | $2.35\times10^{-3}$ | 22  | 2   | -3.97 | 0.12  | 22  |
| 10:35 | 22.12 | 49  | -5.18 | 221 | 50  | -5.21  | $5.61\times10^{-6}$ | 220 | 21  | -2.97 | 5.40  | 218 |
| 10:40 | 29.87 | 55  | -5.76 | 311 | 4   | -0.01  | $4.15\times10^{0}$  | 274 | 9   | -0.54 | 15.64 | 273 |
| 10:45 | 37.38 | 98  | -6.33 | 173 | 40  | -3.97  | $2.10\times10^{0}$  | 131 | 19  | -0.98 | 17.76 | 132 |
| 10:50 | 41.66 | 120 | -6.64 | 193 | 100 | -5.76  | $1.01\times10^{0}$  | 157 | 80  | -4.63 | 7.35  | 152 |
| 10:55 | 42.07 | 114 | -6.86 | 200 | 132 | -6.89  | $3.80\times10^{-1}$ | 146 | 86  | -5.10 | 7.38  | 157 |
| 11:00 | 39.00 | 101 | -6.98 | 200 | 142 | -7.57  | $4.78\times10^{-3}$ | 138 | 79  | -5.22 | 7.95  | 158 |
| 11:10 | 38.43 | 119 | -7.31 | 290 | 189 | -7.96  | $6.40\times10^{-4}$ | 222 | 109 | -5.93 | 7.06  | 253 |
| 11:20 | 35.03 | 118 | -7.54 | 261 | 186 | -8.43  | $3.91\times10^{-3}$ | 186 | 116 | -5.64 | 11.73 | 203 |
| 11:30 | 28.58 | 98  | -7.76 | 187 | 179 | -8.68  | $2.71\times10^{-9}$ | 117 | 105 | -6.60 | 7.62  | 160 |
| 11:40 | 26.00 | 96  | -7.67 | 233 | 191 | -8.76  | $8.60\times10^{-3}$ | 164 | 104 | -5.53 | 14.68 | 180 |
| 11:50 | 23.98 | 109 | -7.91 | 161 | 241 | -9.19  | $4.46\times10^{-6}$ | 100 | 129 | -6.48 | 11.00 | 127 |
| 12:00 | 21.96 | 106 | -8.00 | 165 | 239 | -9.41  | $4.61\times10^{-4}$ | 107 | 139 | -5.37 | 24.29 | 108 |
| 12:10 | 19.46 | 89  | -7.95 | 158 | 204 | -9.46  | $3.48\times10^{-6}$ | 96  | 108 | -6.58 | 11.90 | 135 |
| 12:20 | 17.36 | 86  | -8.14 | 116 | 211 | -10.00 | $1.26\times10^{-3}$ | 63  | 116 | -5.90 | 22.60 | 66  |
| 12:30 | 17.44 | 89  | -8.27 | 145 | 212 | -9.77  | $1.34\times10^{-3}$ | 95  | 86  | -6.67 | 16.64 | 125 |
| 12:40 | 14.70 | 61  | -7.82 | 139 | 136 | -9.09  | $1.75\times10^{-9}$ | 95  | 68  | -5.72 | 16.27 | 109 |
| 12:50 | 15.03 | 67  | -7.97 | 99  | 103 | -9.33  | $4.02\times10^{-8}$ | 74  | 80  | -6.36 | 13.36 | 83  |
| 13:00 | 14.78 | 66  | -7.83 | 101 | 147 | -9.85  | $5.49\times10^{-2}$ | 75  | 79  | -5.82 | 15.67 | 90  |
| 13:10 | 12.36 | 53  | -7.95 | 90  | 117 | -9.35  | $1.20\times10^{-5}$ | 63  | 62  | -6.28 | 13.70 | 83  |
| 13:20 | 12.20 | 59  | -8.03 | 81  | 111 | -9.29  | $8.24\times10^{-9}$ | 62  | 54  | -6.16 | 13.02 | 81  |
| 13:30 | 11.07 | 49  | -7.81 | 65  | 89  | -8.76  | $7.72\times10^{-6}$ | 56  | 50  | -6.16 | 10.45 | 84  |
| 13:40 | 10.58 | 49  | -8.05 | 76  | 90  | -8.78  | $6.31\times10^{-1}$ | 71  | 54  | -5.62 | 18.45 | 64  |
| 13:50 | 10.02 | 48  | -7.97 | 54  | 50  | -6.21  | $5.21\times10^{0}$  | 49  | 50  | -7.57 | 2.13  | 53  |
| 14:00 | 10.02 | 44  | -8.00 | 53  | 56  | -7.20  | $2.80\times10^{0}$  | 51  | 47  | -5.97 | 15.72 | 48  |

[a] Time (UT) refers to the start of a five minute interval.
[b] Sea-level corrected percentage increases above the pre-event galactic cosmic ray background of the normalization station, SANAE.
[c] Flux (p/cm$^2$.s.sr.GV) at 1 GV summed over the forward steradian.
[d] Spectral slope ($\gamma$).
[e] Best fit weighted sum of squares employing the power law spectral form.
[f] Modifed power law spectral modifier ($\delta\gamma$).
[g] Best fit weighted sum of squares employing the modified power law spectral form.
[h] Ellison & Ramaty spectral modifier ($\delta\gamma$).
[i] Best fit weighted sum of squares employing the modified Ellison & Ramaty (1985) spectral form.

# Tables 3 and 4

TABLE III

VARIABLE MODEL PARAMETERS: SHOCK ACCELERATION

| Time[a] (UT) | $r$[b] | $E_0$[c] (MeV) | WSS[d] |
|---|---|---|---|
| 10:45 | 1.95 ±0.03 | 1872 ±10 | 136 |
| 10:45[e] | 1.95 ±0.02 | 1871 ±07 | 59 |
| 10:50 | 1.83 ±0.02 | 1942 ±03 | 150 |
| 10:55 | 1.81 ±0.03 | 1852 ±05 | 491 |
| 11:00 | 1.84 ±0.04 | 1675 ±69 | 995 |
| 11:40 | 1.75 ±0.04 | 1277 ±54 | 985 |

[a] Time refers to the start of a five minute interval
[b] Shock compression ratio
[c] $e$-folding energy
[d] Weighted sum of squares
[e] Spectral form fitted from 305 to 4000 MeV

TABLE IV

VARIABLE MODEL PARAMETERS: STOCHASTIC ACCELERATION

| Time[a] (UT) | $N$[b] | $\alpha$[c] (s$^{-1}$) | WSS[d] |
|---|---|---|---|
| 10:45 | 626 ± 173 | 0.054 ±0.002 | 345 |
| 10:45[e] | 666 ± 168 | 0.053 ±0.002 | 253 |
| 10:50 | 2379 ± 527 | 0.044 ±0.001 | 268 |
| 10:55 | 2883 ± 410 | 0.041 ±0.001 | 109 |
| 11:00 | 2846 ± 490 | 0.040 ±0.001 | 172 |
| 11:40 | 9730 ± 1651 | 0.030 ±0.001 | 123 |

[a] Time refers to the start of a five minute interval
[b] Normalization factor
[c] Acceleration efficiency
[d] Weighted sum of squares
[e] Spectral form fitted from 305 to 4000 MeV

Figure 1

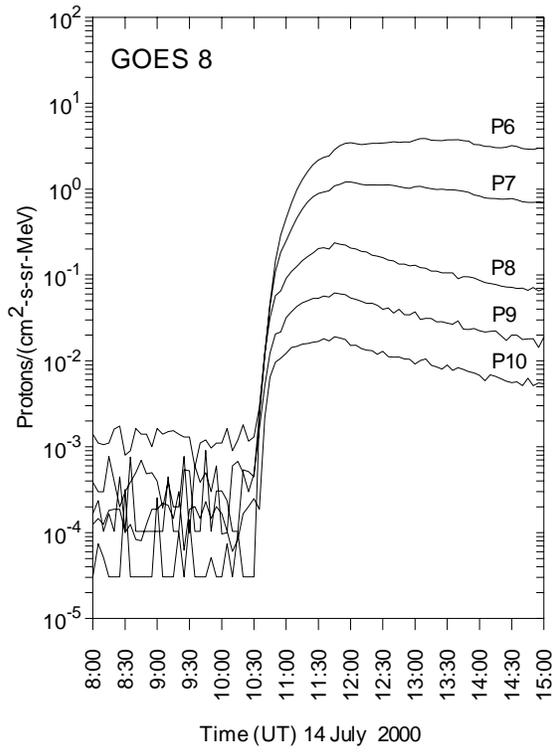

Figure 2

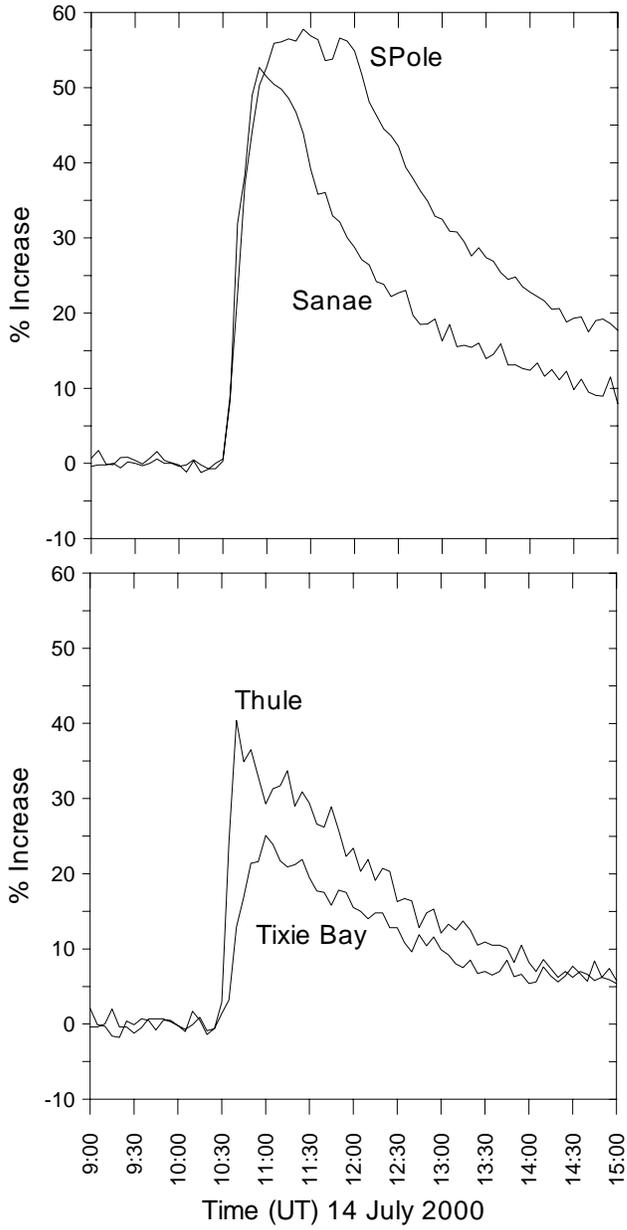

Figure 3

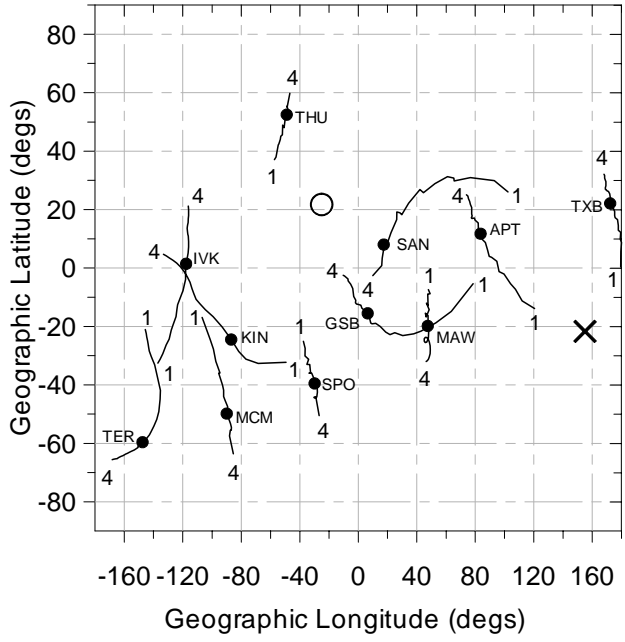

Figure 4

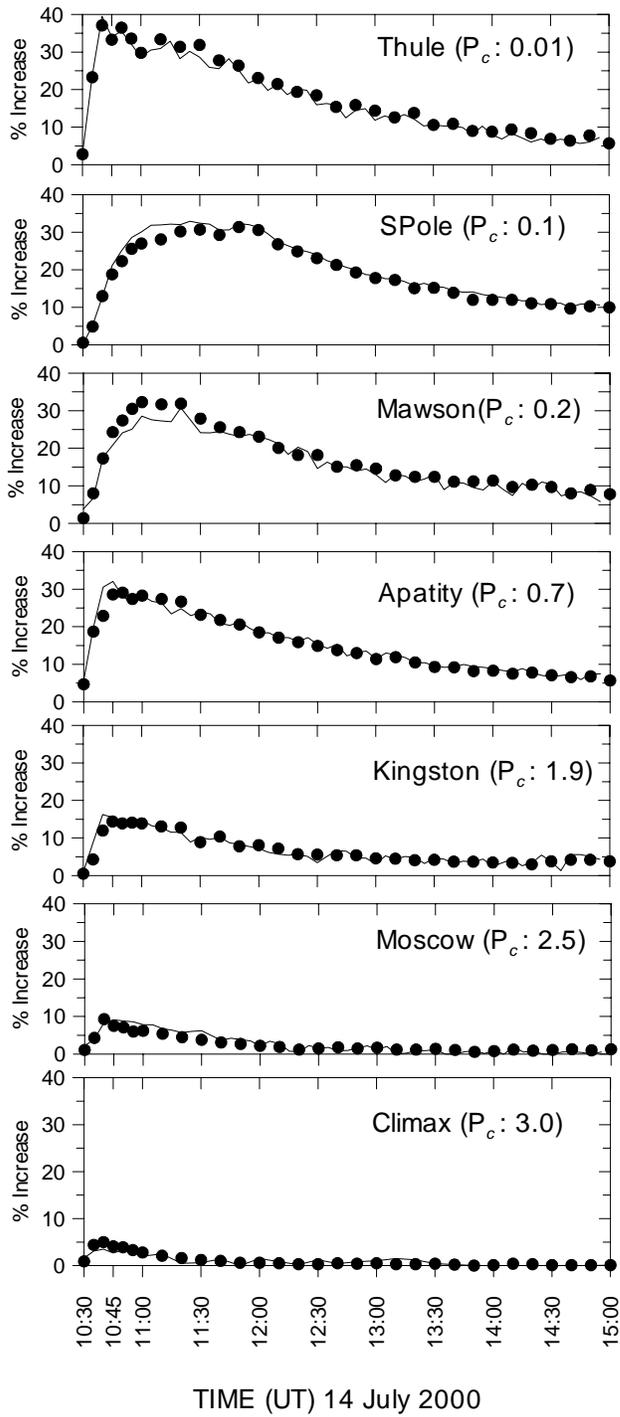

TIME (UT) 14 July 2000

Figure 5

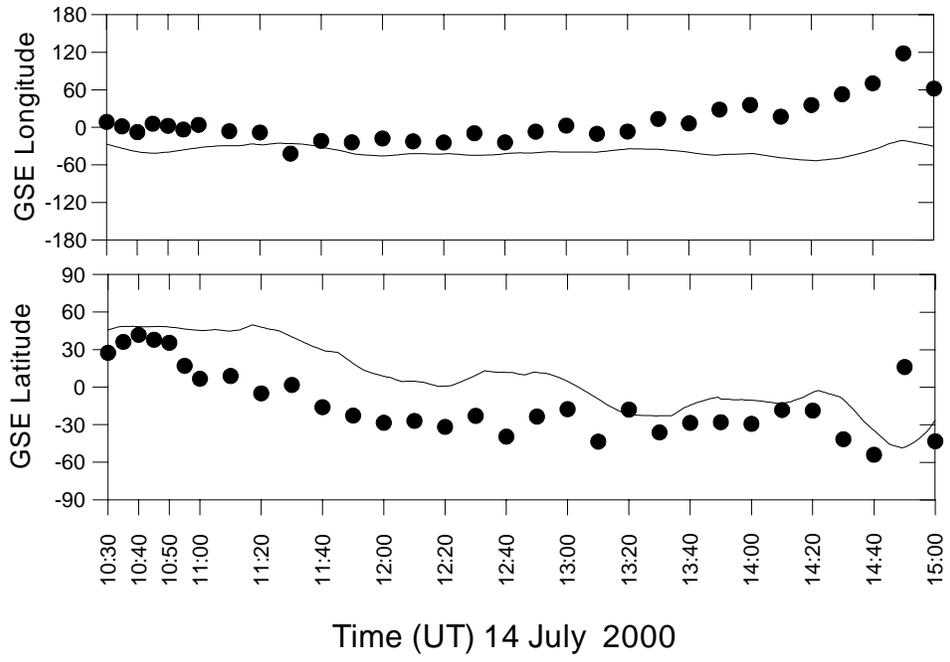

Figure 6

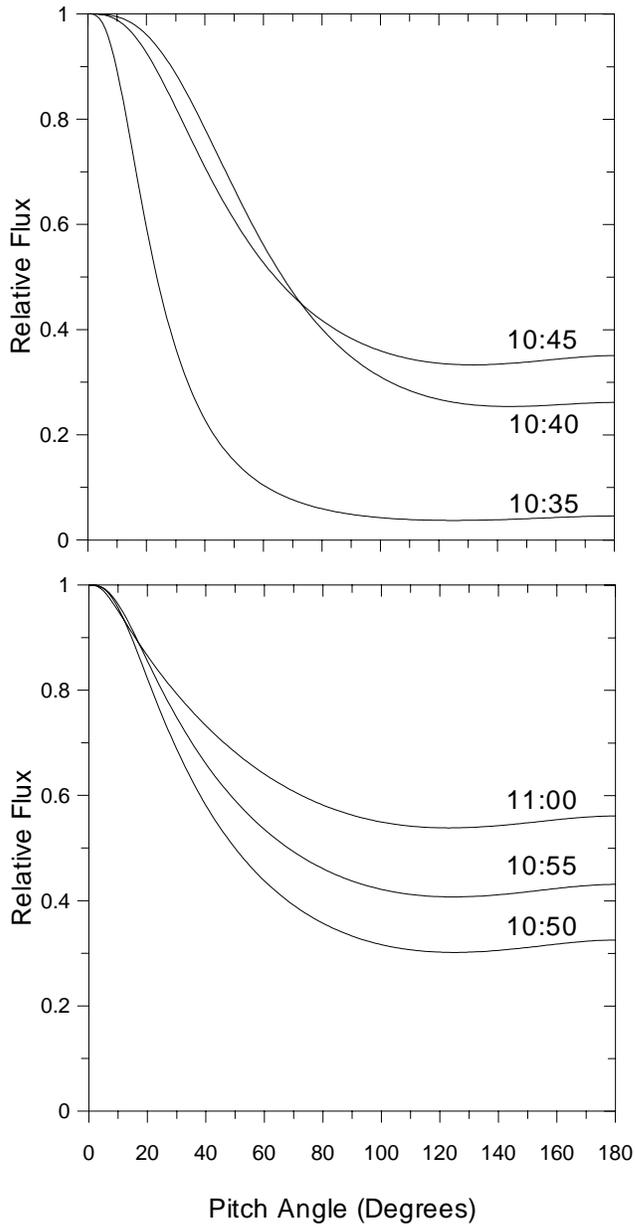

Figure 7

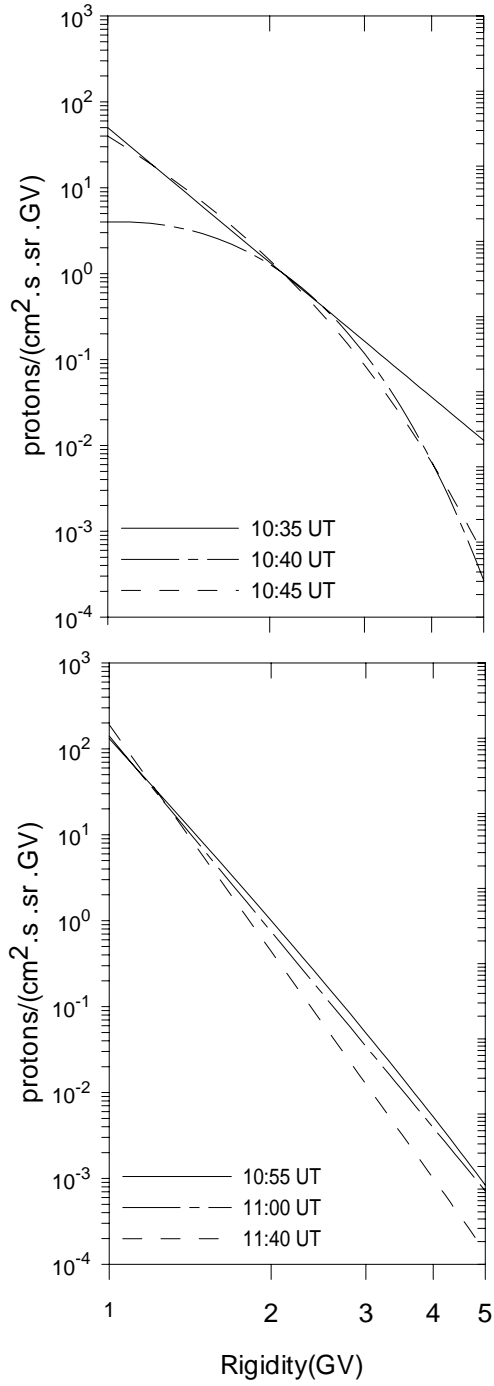

Figure 8

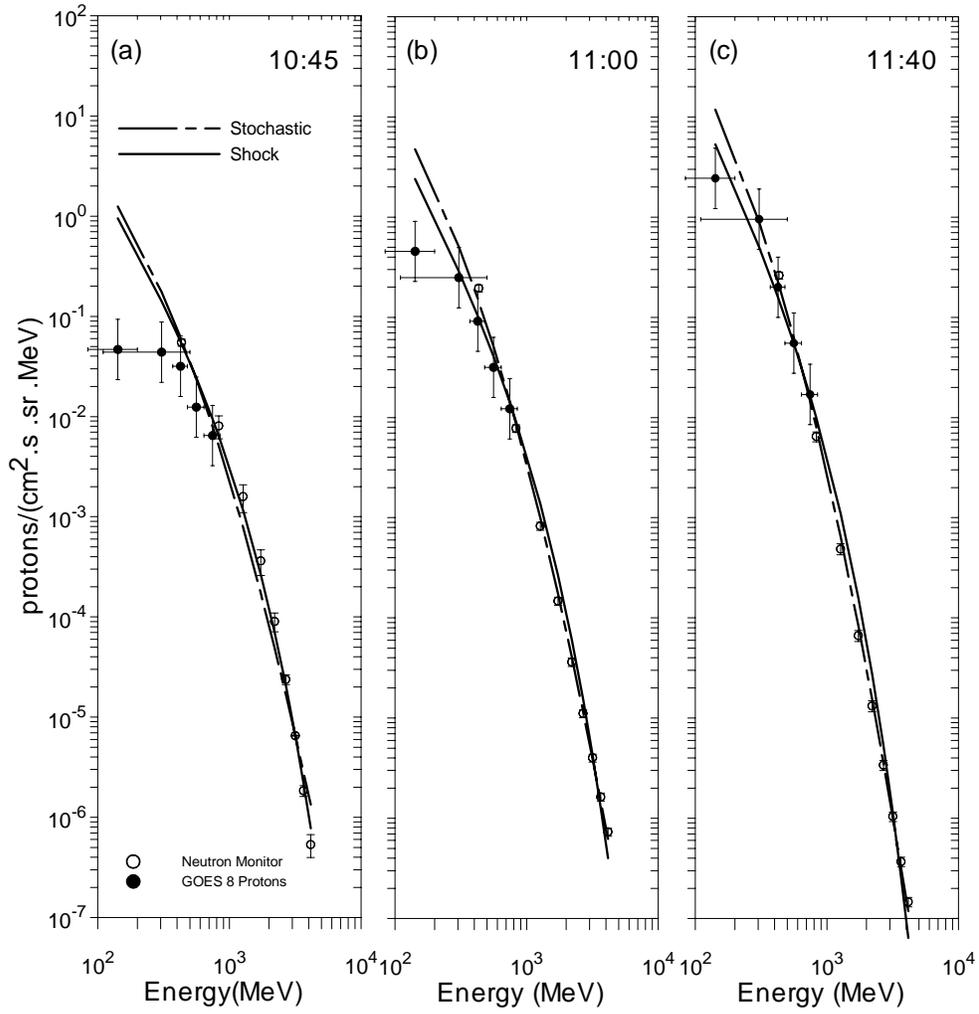

Figure 9

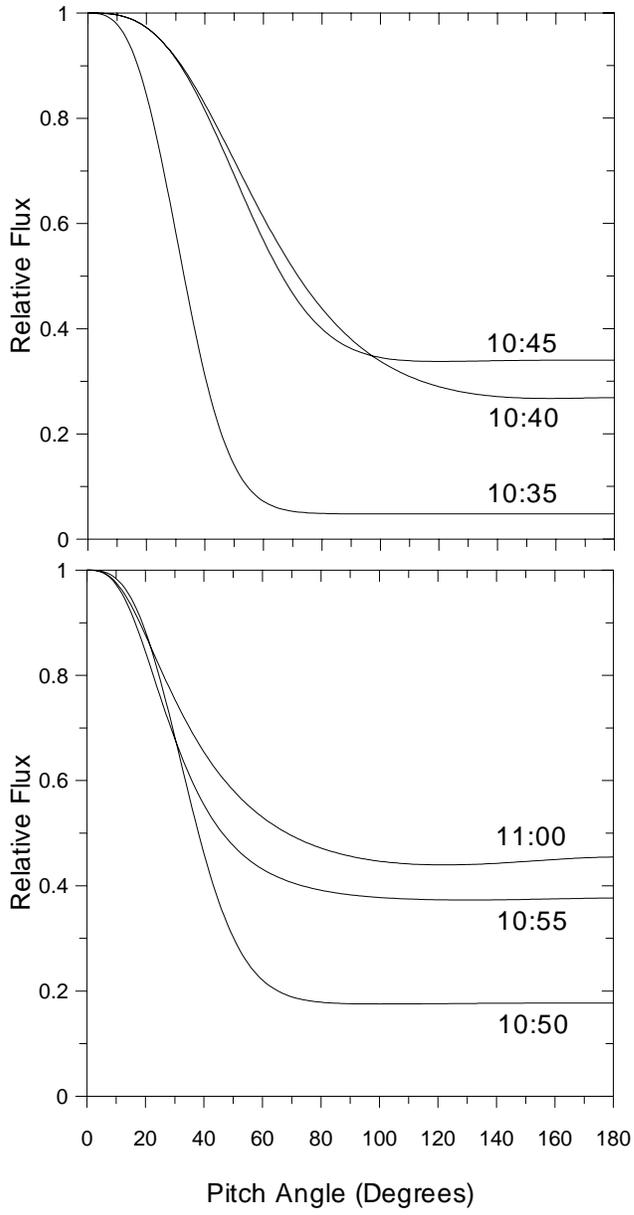